\documentclass[twocolumn,showpacs,preprintnumbers,prb]{revtex4-1}
\usepackage{amssymb}
\usepackage[dvips]{color}
\usepackage{epsfig}
\usepackage{dcolumn}
\usepackage{bm}

\def\G{\Gamma}

\def\D{\Delta}
\def\e{\epsilon}

\def\w{\omega}

\def\s{\sigma}

\begin{document}

\title{Enhanced photon-assisted spin transport in a quantum dot attached to
ferromagnetic leads}

\author{Fabr\'icio M. Souza, Thiago L. Carrara, and E. Vernek}
\affiliation{Instituto de F\'isica, Universidade Federal de Uberl\^andia,
38400-902, Uberl\^{a}ndia,
MG, Brazil} 

\begin{abstract}
We investigate real-time dynamics of spin-polarized current in a quantum dot
coupled to ferromagnetic leads in both parallel and antiparallel alignments. While an external bias voltage is taken
constant in time, a gate terminal, capacitively coupled to the quantum dot, introduces a periodic modulation of the
dot level. Using non equilibrium Green's function technique we find that spin polarized
electrons can tunnel through the system via additional photon-assisted
transmission channels. Owing to a Zeeman splitting of the dot level, it is possible to select a
particular spin component to be photon-transferred from the left to the right terminal, with spin
dependent current peaks arising at different gate frequencies. The ferromagnetic
electrodes {\it enhance} or {\it suppress} the spin transport depending {upon} the leads
magnetization alignment. The tunnel magnetoresistance also attains negative
values due to a photon-assisted inversion of the spin-valve effect.
\end{abstract}

\volumeyear{year} \volumenumber{number} \issuenumber{number}
\eid{identifier}
\date[Date: ]{\today}
\maketitle

\section{Introduction}

Time-dependent transport in quantum dot system (QDs) has received significant
attention due to a variety of new quantum physical phenomena emerging in transient time
scale.\cite{nsw93} A few examples encompass charge pump\cite{ljg90,lpk91} and photon-assisted tunneling 
transport.\cite{cb94,lpk94_1,lpk94_2,bjk95,gp04} {For instance, a double dot junction sandwiched by leads can be used to
pump electrons uphill from a lead with lower chemical potential to a lead with higher chemical
potential, in contradiction to the usual dc-regime.\cite{cas96} This was achieved by applying
a sinusoidal  gate voltage  among the dots.}
Photon-assisted tunneling can occur when an oscillating gate potential or laser field is applied
in a QD or a metallic central island
{coupled to source and drain
terminals}.\cite{cb94,lpk94_1,lpk94_2,bjk95,gp04,cas96,afa09}
Time-dependent regime also leads to {zero-bias} charge or spin pumping
when a {minimum set of} two parameters of the system (e.g., gate
potential and tunneling rate in a QD system) are time modulated independently.
{This is the case, for instance, of the non-adiabatic charge and spin
pumping through
interacting quantum dots\cite{fc09} and  quantum  pumping
in graphene-based structures.\cite{rz09,ep09,gmmw10,map11}}

Transient charge and spin dynamics in
an interacting QD driven by step pulse or sinusoidal 
gate voltages revealed distinct charge and spin relaxation times.\cite{js10} 
{An exquisite behavior that has been predicted theoretically is the 
self-sustained current oscillations  in a quantum dot system driven
out-of-equilibrium by
a fast switching on of the bias voltage, contrasting to the expected
steady state behavior. This phenomena has been attributed
to dynamical Coulomb blockade.\cite{sk10}}

It is in the fascinating area of spintronics\cite{spintronics}
that time-dependent quantum transport reveals its prolific potentiality in
producing spin polarized currents. For instance, a double dot
structure driven by ac-field in the presence
of magnetic field turn out to be a robust spin {filtering and
pumping device}.\cite{ec05,rs06} 
By applying oscillating gates (radio frequency) in an open quantum dot in the
presence of Zeeman
field, an adiabatic spin pump was generated.\cite{erm02,ta03,skw03} The current
ringing\cite{nsw93}
that arises in a quantum dot system when a bias voltage is suddenly switched on
develops spin
dependent beats when the dot level is Zeeman split.\cite{fms07_1,ep08} Coherent
quantum beats in the
current, spin current and tunnel magnetoresistance of two dots coupled to three
ferromagnetic leads
were also reported recently.\cite{pt10} Additionally, spin spikes take place
when a bias voltage is
abruptly turned off in a system of a QD attached to ferromagnetic
leads.\cite{fms07_2,fms09}

{The study of quantum transport of spin polarized electrons in the
presence of time varying fields was greatly motivated by the
development of experimental techniques. These techniques allow for the  coherent
control 
of the complete} dynamics (initialization-manipulation-read out)
of single electron spins in quantum dots.\cite{rh08} Particularly, some of 
{those} {used to coherently manipulate spin states are based on} time
dependent gate voltages.\cite{jme04,rh05,acj05,jrp05,fhlk06,cb10}

In the present work we consider a single level quantum dot coupled to
{left and right ferromagnetic leads}
in the presence of a static bias voltage and sinusoidal gate voltage.
{The} oscillating gate potential introduces 
additional photon-assisted conduction channels that can be tunned via a dc-gate
 field to {lie within the conduction window of the system. In the
presence of Zeeman splitting, produced by an applied magnetic field, the
 contribution from the photon-assisted channels becomes different for
spins up and down, resulting in photon-assisted spin polarized currents.
It is worth mentioning that this effect takes place even in
the absence of ferromagnetic leads. However,  when the leads are ferromagnetic
and parallel aligned, the resonant current peaks are amplified for one spin
component and suppressed for the other. Thus the photon-assisted
current-polarization is enhanced.} We also calculate the tunnel
magnetoresistance (TMR) as a function of the gate frequency, {which exhibits a
variety
of peaks and dips, having even a changed of sign, depending on the gate
frequency.}  

The paper is organized as {follows: in} Sec.~\ref{modelsec} we present the
theoretical model and describe
the formulation based on nonequilibrium {Green's} function technique and
in Sec.~\ref{results} we show and discuss the numerical
results. Finally, in Sec.~\ref{conclusions} we present our concluding remarks.

\begin{figure}[t]
\par
\begin{center}
\epsfig{file=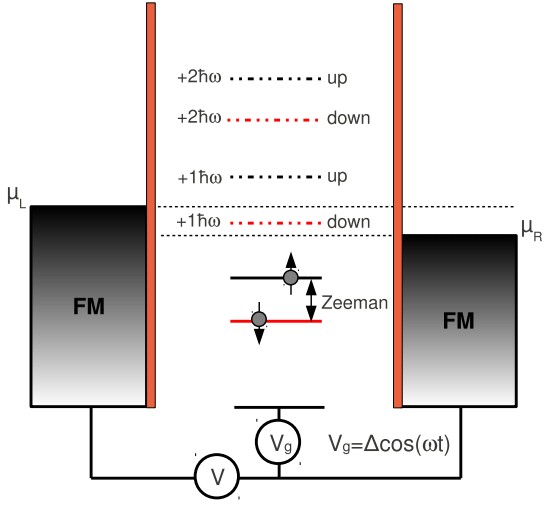, width=0.45\textwidth}
\end{center}
\caption{(color online) Energy diagram for the system considered. A quantum dot
is coupled to
a left and to a right ferromagnetic electron
reservoirs via tunneling barriers. The dot level is Zeeman split. A capacitively
coupled gate
terminal
introduces a periodic perturbation of the dot level. This modulation induces
additional
photon-assisted
channels (dashed lines) for spin polarized transport.}\label{fig1}
\end{figure}

\section{Model and theoretical formulation}
\label{modelsec}

For concreteness, the energy profile of our system is illustrated in
Fig.~\ref{fig1} and is described by the
Hamiltonian, $H=H_L+H_R+H_D(t)+H_T$, where
\begin{equation}
H_{L(R)}=\sum_{\mathbf{k} \s} \e_{\mathbf{k} \s L(R)}
c_{\mathbf{k} \s L(R)}^\dagger c_{\mathbf{k} \s L(R)},
\end{equation}
describes the free electrons in the ($L$) or the right ($R$) lead, in which
 $c_{\mathbf{k} \s L(R)}$ [$c_{\mathbf{k} \s
L(R)}^\dagger$] is the operator that annihilates [creates] an
electrons in the lead $L(R)$ with momentum $\mathbf{k}$, spin $\s$ and
energy $\e_{\mathbf{k} \s L(R)}$. 
We consider a static source-drain
applied voltage ($e V_{SD} = \mu_L-\mu_R$) which drives the system out of
equilibrium, breaking
the left/right symmetry of the Hamiltonian. The time dependence of our
Hamiltonian is fully accounted via the dot Hamiltonian,
\begin{equation}
H_D(t)=\sum_{\s} \e_\s(t) d_\s^\dagger d_\s,
\end{equation}
where $\e_\s(t)=\e_d(t)+\s E_Z/2$, with $\e_d(t)$ being the time-dependent
dot level and $E_Z$ a Zeeman splitting of the dot level due to an external
magnetic field. Here we use $\s=+$ and $\s=-$ for spins up and down,
respectively.
The operator $d_\s$ ($d_\s^\dagger$) annihilates (creates) one electron with
spin $\s$ and energy $\e_\s(t)$ in the dot. 
In practice the time dependence in the dot level is controlled by an
oscillating gate voltage $V_g(t)$, such that $\e_d(t) = \e_0 + e V_g(t)$, where
$\e_0$ is the dc component of the energy and $e V_g(t) = \Delta \mathrm{cos} (\w
t)$ oscillates with amplitude $\D$ and frequency $\omega$. Finally
\begin{equation}
H_T=\sum_{\mathbf{k} \s \eta} (V c_{\mathbf{k} \eta}^\dagger d_\s
+V^* d_\s^\dagger c_{\mathbf{k} \s \eta}),
\end{equation}
describes the tunnel coupling between the leads and the dot, with a constant
coupling strength $V$ and allows for current to flow across the QD.

To calculate the time dependent spin polarized current we employ 
the Keldysh Green's function formalism\cite{hh96} that allows for an
appropriate approach to our nonequilibrium time-dependent situation. 
Starting from the current definition $I_\s^\eta(t)=-e \langle \dot{N}_\s 
\rangle = -i e \langle [H,N_\s] \rangle$, where $N_\s$
is the total number of particle operator for spin $\s$ (here we take $\hbar=1$),
the current can be
written as\cite{apj94}
\begin{equation}\label{current}
 I^\eta_\s(t) = 2 e \mathrm{Re} \left\{ \sum_{k} V
G_{\s,k\s\eta}^<(t,t)\right\},
\end{equation}
where $G_{\s,k\s\eta}^<(t,t')=i \langle c_{k \s \eta}^\dagger (t') d_\s (t)
\rangle$. Using the equation of motion technique and taking analytical
continuation\cite{hh96} to 
obtain $G_{\s,k\s\eta}^<(t,t')$ one finds to the current the following
\begin{eqnarray}
 I^\eta_\s(t) &=& -2e \Gamma_{\eta}^\s \mathrm{Im} \Big\{\int \frac{d\e}{2\pi} 
\int_{-\infty}^t dt_1 e^{-i\e(t_1-t)} \nonumber \\  && \phantom{xxxxxx}\times 
[G_{\s\s}^r(t,t_1) f_\eta(\e) + G_{\s\s}^<(t,t_1)]\Big\},
\end{eqnarray}
where $f_\eta(\e)$ is the Fermi distribution function of the  $\eta$-th lead,
and 
$\Gamma^{\eta}_\s=2\pi |V|^2 \rho^{\eta}_\s$ gives the tunneling rate
between lead $\eta$ and dot for spin component $\s$.
$\rho^\eta_\s$ is the density of states for spin $\s$ in lead $\eta$.  In the
present model we assume constant density of states (wide-band limit). The
ferromagnetism of the electrodes is modeled by considering $\G_\s^\eta=\G_0
(1 \pm p_\eta)$ where $+$ ($-$) stands for spin up (down), $p_\eta$ is the
polarization of lead
$\eta$-th\cite{wr01,fms04} and $\G_0$ the tunneling rate strength.
{The quantity
$\G_0$
is fixed along the paper, so all the other energies will be expressed in units
of $\G_0$.
We consider both parallel (P) and antiparallel (AP) alignments of the lead
polarizations. In the 
P case we assume majority down population in both leads, while in the AP
configuration we take
majority down population in the left lead and majority up population in the
right lead. In terms
of the parameters $p_\eta$ we have $p_L=p_R=p=-0.4$ for the P and
$p_L=-p_R=p=-0.4$ for the AP
case.}\cite{commentnoncollinear}
\begin{figure}
\par
\begin{center}
\epsfig{file=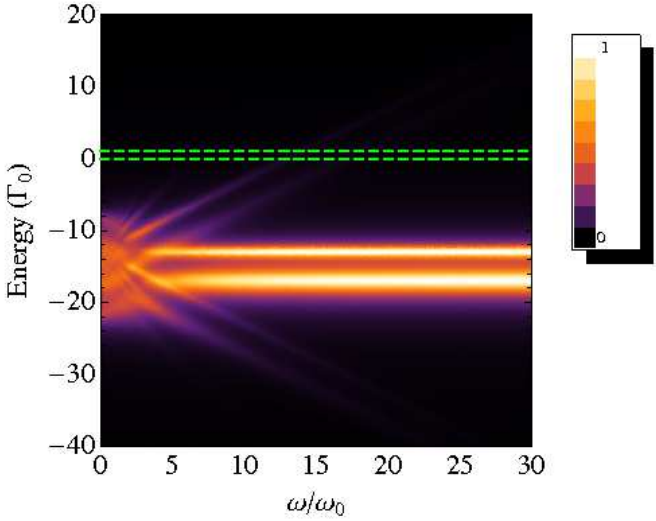, width=0.5\textwidth}
\end{center}
\caption{(color online) Color map of the total transmission coefficient 
$T(\e,\w) =
T_\uparrow (\e,\w) + T_\downarrow (\e,\w)$ as a function of
frequency and energy in the parallel alignment ($p_L=p_R=-0.4$). For increasing
$\w$, $T(\e,\w)$
develops additional 
photon-assisted peaks that allows off-resonant spin transport.
The main two central peaks correspond to the Zeeman split levels $\e_\uparrow^0
= \e_0 + E_Z/2$  and $\e_\downarrow^0 = \e_0 - E_Z/2$.
The satellite peaks are given by $\e_\uparrow^{(n)} = \e_0 + E_Z/2 \pm n \w$ and
$\e_\downarrow^{(n)} = \e_0 - E_Z/2 \pm n \w$, with $n=1,2,3,...$. For
increasing
$\w$, the satellite peaks tend to vanish and the system recovers its original
two
levels $\e_\uparrow^0$ and $\e_\downarrow^0$. The horizontal dashed lines
delimit the conduction
window [$\mu_R,\mu_L$].
Units: Energy in units of $\G_0$ and $\w_0=\G_0 / \hbar$.
Parameters: $\e_0=-15\G_0$, $E_Z=4\G_0$, $\D=5\G_0$, $\mu_L=1\G_0$, $\mu_R=0$.}
\label{fig2}
\end{figure}
Taking the time average of the current we find
\begin{equation}\label{I1}
 \langle I^L_\s(t) \rangle = -2 e \frac{\G_\s^L \G_\s^R}{\G_\s^L + \G_\s^R} \int
\frac{d\e}{2\pi}
[f_L(\e)-f_R(\e)] 
  \mathrm{Im} \langle A_\s(\e,t) \rangle,
\end{equation}
where
\begin{equation}
 \langle A_\s(\e,t) \rangle = \sum_{n=-\infty}^{\infty} J_n^2(\frac{\D}{\w})
g_{n,\s}^R(\e,\w),
\end{equation}
{{with $J_n$ being the n-th order} Bessel function and}
$g_{n,\s}^R(\e,\w)=[\e-\e_\s^0-n\w+i\frac{\G_\s^L+\G_\s^R}{2}]^{-1}$. 
Here we used the fact that $\e_\s(t)=\e_\s^0+\D \cos(\w t)$, with
$\e_\s^0=\e_0+\s E_Z/2$.

Substituting this result into Eq. (\ref{I1}), the current can be written in its 
Landauer form\cite{commentreviewplatero}
\begin{eqnarray}\label{Ifinal1}
\langle I_\s^L(t) \rangle &=&  e \int \frac{d\e}{2\pi} T_\s(\e)
[f_L(\e)-f_R(\e)].\nonumber\\
\end{eqnarray}
Here we define
\begin{equation}\label{Ts}
 T_\s(\e)= \G_\s^L \G_\s^R \sum_{n=-\infty}^{\infty}
\frac{J_n^2(\frac{\D}{\w})}{(\e-\e_\s^{(n)})^2+(\frac{\G_\s}{2})^2},
\end{equation}
where $\e_\s^{(n)}=\e_\s^0+n\w$. Eq. (\ref{Ts}) shows that the harmonic
modulation of the dot
level yields to photon-assisted peaks in the transmission
coefficient.\cite{gp04} 
In addition to this, here we have the spin splitting of these peaks and the
ferromagnetic leads,
that results in an enhanced spin photon-assisted transport.

A further simplification can be made in Eq. (\ref{Ifinal1}) by considering the
low  temperature regime, where the Fermi functions are
approximated by step functions. In this regime, the integral in Eq.
(\ref{Ifinal1})  is carried out in the range $[\mu_L,\mu_R]$,
thus resulting in 
\begin{equation}
 \langle I_\s^L \rangle =I_\s^0 \Phi_\s,
\end{equation}
where $I_\s^0=e\frac{\G_\s^L \G_\s^R}{\G_\s^L + \G_\s^R}$ is the resonant
current  without modulated gate voltage and
\begin{equation}\label{phi}
 \Phi_\s= \sum_{n=-\infty}^\infty J_n^2(\frac{\D}{\w}) [\Theta_n^{\s
L}(\w)-\Theta_n^{\s
R}(\w)]/\pi,
\end{equation}
with $\Theta_n^{\s\eta}(\w)=\arctan [2(\mu_\eta-\e_\s^0-n \w)/\G_\s]$. In what 
follows we present our numerical results to the spin polarized transport.

\section{Numerical results}
\label{results}
\begin{figure}
\par
\begin{center}
\epsfig{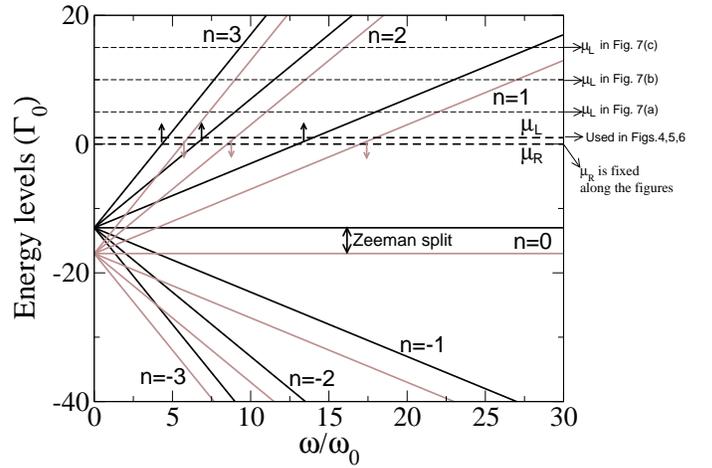}
\end{center}
\caption{(color online) Multiplet structure developed in the presence of an
oscillating gate
frequency.  The black lines correspond to spin up while
the gray lines to spin down. The levels are shifted linearly with the gate
frequency,  following $\e_\s^{(n)}=\e_\s^0 \pm n \w$, $n=1,2,3,...$
The up and down levels are Zeeman split. The horizontal dashed lines correspond
to  the left ($\mu_L$) and to the right ($\mu_R$) chemical
potentials. The channels $\e_\uparrow^{(n)}$ and $\e_\downarrow^{(n)}$ attain 
resonance within the conduction window
$[\mu_L,\mu_R]$ for certain frequencies, which differ for each spin component. 
Units: Energy levels in units of $\G_0$ and $\w_0=\G_0 / \hbar$.
Parameters: $\e_0=-15\G_0$, $E_Z=4\G_0$, $\D=5\G_0$, $\mu_L=1\G_0$, $\mu_R=0$.}
\label{fig3}
\end{figure}

Figure \ref{fig2} shows the sum $T=T_\uparrow+T_\downarrow$ as a function of
$\w$ and energy
in the case of polarized leads with parallel magnetizations. As $\w$
increases,  a multiplet structure takes place in the 
transmission coefficient [Eq. (\ref{Ts})]. The two central peaks in $T(\e,\w)$ 
correspond to $\e_\uparrow^0$ and $\e_\downarrow^0$, while the lateral peaks are
related to 
$\e_\s^0 \pm n\w$. Due to the Zeeman splitting, the whole pattern for
$T_\uparrow$ 
is shifted upward while $T_\downarrow$ is moved downward. The highest of the
peaks  are strongly affected by the frequency.
For the n-th peak its amplitude is given by $4 \G_\s^L \G_\s^R J_n^2 (\Delta/\w)
 /\Gamma_\s^2$. For
sufficiently large $\w$, the additional photon-assisted peaks
are suppressed, remaining only the two central peaks. The broadening difference
for up 
and down spin channels comes from the ferromagnetism of the electrodes that are
parallel
aligned, with majority down population in both sides ($p_L=p_R=-0.4$).
{This gives
rise to narrower spin up peaks than spin down ones. In the case of
parallel aligned leads with majority up populations,
we have basically the same structure, {but with an inversion of the peak
widths, with spin
up peaks now becoming broader.}}

\begin{figure}
\par
\begin{center}
\epsfig{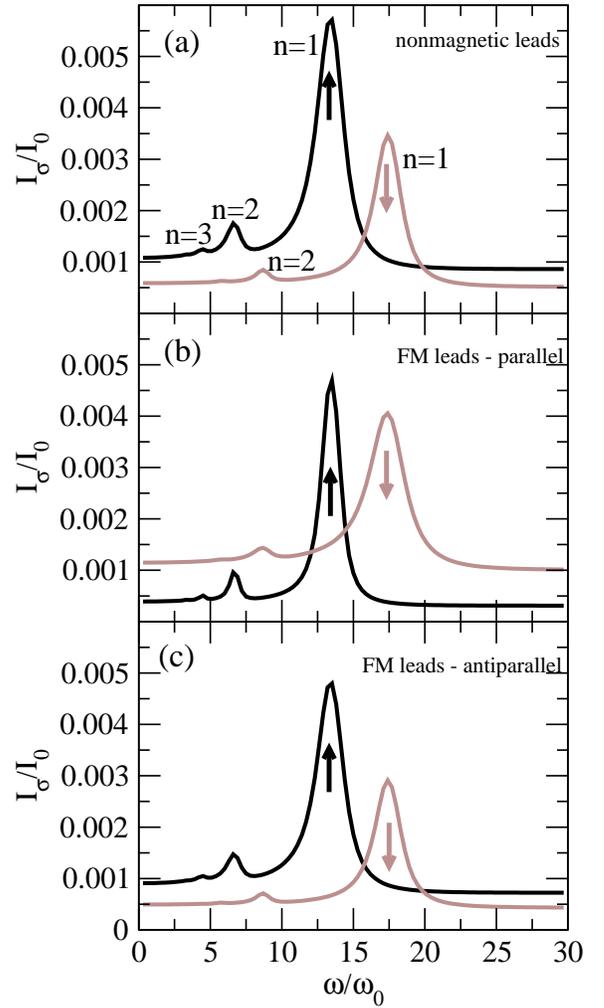}
\end{center}
\caption{(color online) Spin resolved currents against gate frequency for leads
(a)
 nonmagnetic and (b)-(c) ferromagnetic (black lines for spin up and gray lines
for spin down).
In panels (b) and (c) we show the parallel and antiparallel alignments, 
respectively. Both up and down currents 
show peaks corresponding to the crossing of $\e_\uparrow^{(n)}$ and
 $\e_\downarrow^{(n)}$ illustrated in Fig. (\ref{fig3}).
The highest peak for each spin component comes from the resonance of the levels
$\e_\uparrow^{(1)}$
and $\e_\downarrow^{(1)}$,
withing the conduction window. In the parallel alignment the majority down
population in both 
leads turns into an amplification of the down current. In the antiparallel
case, though, the currents are very similar to the nonmagnetic case. 
Units: $I_0=e\G_0/\hbar$ and $\w_0=\G_0/\hbar$.
Parameters: $\e_0=-15\G_0$, $E_Z=4\G_0$, $\D=5\G_0$, $\mu_L=1\G_0$, $\mu_R=0$,
$p=-0.4$.}
\label{fig4}
\end{figure}

Spin polarized transport can arise depending upon the position of the
peaks of $T_\s(\e,\w)$ with respect to the conduction window. 
Fig. (\ref{fig3}) shows the channels $\e_\uparrow^0 + n \w$ and $\e_\downarrow^0
+  n \w$ for $n=0,\pm 1,\pm 2, \pm 3$ as black and gray lines, respectively.
The left and right chemical potentials are indicated by the horizontal dashed
lines. A net electron transport from the left to the right lead
can take place whenever a channel $\e_\s^0 + n\w$ attains the conduction window
(CW)
interval $[\mu_L,\mu_R]$.  Due to the Zeeman splitting, each spin
component crosses $\mu_L$ or $\mu_R$ at different frequencies, thus resulting in
a frequency  selective spin transfer between the leads.
In Fig. (\ref{fig3}) we indicate by up and down arrows the corresponding
crossing of the
CW for spins $\uparrow$ and $\downarrow$, respectively. In the present study
we focus on the off-resonant regime, where the dot levels $\e_\uparrow^0$ and
$\e_\downarrow^0$
are below the CW. In this case only photon-assisted electrons can tunnel through
the system. In order to match this condition we adopt to the numerical
parameters the following
values:
$\e_0=-15\G_0$, $E_Z=4\G_0$, $\D=5\G_0$, $\mu_L=1\G_0$ and $\mu_R=0$. 
Later on we will also look at distinct parameters in order to explore the
robustness of our main
results. 
In experiments we find typically $\G_0 \sim 100\mu
eV$.\cite{dgg98_1,dgg98_2,fs99} So to the
parameters assumed we have $E_Z \sim 400 \mu eV$.
This Zeeman energy split is reasonable for semiconductor quantum dots in the
presence of magnetic
fields $\sim$ 1-10 T.\cite{rh03}
Additionally, {for these values} we find $\w_0 = \frac{\G_0}{\hbar}
\sim 150$ GHz. So the present theoretical effects could
be observed for gate frequencies around 1.5 THz [$\w \sim 10 \w_0$, see Fig.
(\ref{fig4})].\cite{thzelectronics} 
Alternatively, if $\G_0$ is reduced to  $\sim 1 \mu$eV,\cite{sami,km07} we
obtain gate frequencies
around $\w \sim 10 \w_0 \sim 15$ GHz, 
which is quite feasible experimentally.\cite{he10} {Our currents
will be given in
units of $I_0=e \G_0/\hbar$,
which is in the range $I_0 \sim$ 0.24nA-24nA for $\G_0 \sim 1 \mu$eV - 100
$\mu$eV. Since our spin
resolved photon-assisted currents
are typically $\sim 10^{-3}  I_0$, we have pA currents, which could be measured
with picoampere
{measurement technologies}.}

Comparing Fig. (\ref{fig3}) to Fig. (\ref{fig2}) one can note that even though
$n=3$ and $n=2$ attain resonance withing $[\mu_L,\mu_R]$, their corresponding
transmission amplitude are very low, which makes the transport weak via those
channels. In contrast,
the $n=1$ channels,
for up and down spins, have a higher transmission amplitude, which makes the
spin transfer via these
channels
more appreciable. 

Fig. (\ref{fig4}) shows the up and down components of the current against gate
frequency. Three
cases are considered:
(a) nonmagnetic leads, ferromagnetic leads in the (b) parallel and (c)
antiparallel alignments. In
all the three cases
two major peaks are found ($n=1$). Satellite peaks for high order channels
($n=2,3$) are also seen.
Each peak emerges whenever a channel $\e_\s^0+n \w$ enters
the conduction window [indicated
by $\uparrow$ and $\downarrow$ arrows in Fig. (\ref{fig3})]. Due to the Zeeman
splitting, the resonance for spin up arises in lower frequencies than that for
spin down.
Additionally, the interplay
between Zeeman splitting and the amplitude of the transmission coefficient
results into a higher
peak for spin up
than for spin down in the case of nonmagnetic leads.\cite{commentotherwork}
Further amplification of the spin down current peak is observed when the leads
are made ferromagnetic. In Fig. \ref{fig4}(b) we present $I_\uparrow$ and
$I_\downarrow$ for leads
parallel aligned, with a majority down population in both sides. This means that
we assume for the
polarization
 parameter a negative value, with $p_L=p_R=p=-0.4$. This implies that
$\G_\downarrow^{L,R} > \G_\uparrow^{L,R}$ ($\rho_\downarrow^{L,R} >
\rho_\uparrow^{L,R}$), which
favors more the spin down electrons
to tunnel through the system, thus increasing the spin down current peak. In the
antiparallel case,
where we have a majority
down population in the left lead and a majority up population in the right lead 
($\G_\downarrow^L > \G_\uparrow^L$ and $\G_\downarrow^R < \G_\uparrow^R$ for
$p_L=-p_R=p=-0.4$), the
incoming and outgoing rates compensate each
other. This {results} into equal weights for both up and down currents,
so the
current remains almost
the same
compared to the nonmagnetic case.

\begin{figure}
\par
\begin{center}
\epsfig{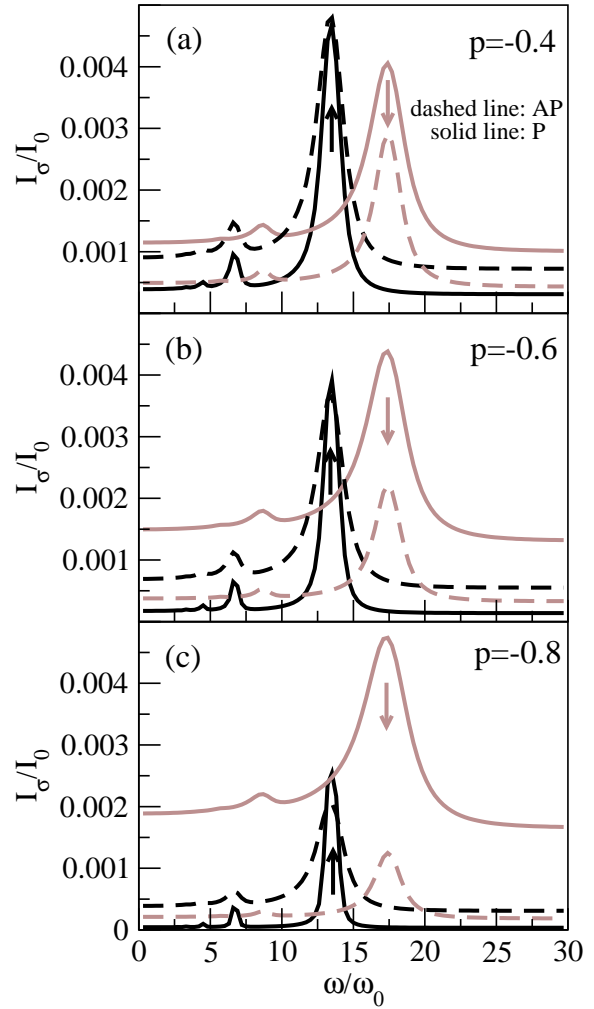}
\end{center}
\caption{(color online) Spin resolved currents against gate frequency for
different polarization $p$,
in both parallel and antiparallel configurations (black lines for spin up and
gray lines for spin
down). In the plots $p$ is negative, which means
that both leads have majority spin down population in the P case and majority
down (up) population
in the
left (right) lead for the AP case. In the P alignment when $p$ increases (in
modulus) the tunneling 
rates between the dot and the leads enlarge for spin down and reduce for spin
up. 
This results in an amplification of $I_\downarrow$ and a suppression of
$I_\uparrow$ as observed.
In the AP alignment both $I_\uparrow$ and $I_\downarrow$ are suppressed as $|p|$
increases.
Units: $I_0=e\G_0/\hbar$ and $\w_0=\G_0/\hbar$.
Parameters: $\e_0=-15\G_0$, $E_Z=4\G_0$, $\D=5\G_0$, $\mu_L=1\G_0$,
$\mu_R=0$.} \label{fig5}
\end{figure}

Fig. (\ref{fig5}) shows how the spin resolved currents evolve when the leads
polarizations are
enlarged
in both P and AP configurations. When $p$ becomes more negative, the spin down
tunneling rates $\G_\downarrow^L$ and $\G_\downarrow^R$ are strengthened, while
$\G_\uparrow^L$ and
$\G_\uparrow^R$ are diminished in the parallel case.
This amplifies the peak for spin down current while suppresses the peak for spin
up, thus making the current more down polarized. 
Eventually, for larger enough $|p|$ the $I_\downarrow$ current dominates over
$I_\uparrow$ for all
gate frequencies,
Conversely, in the antiparallel configuration, as $p$ {becomes} more
negative, both
$I_\uparrow$ and $I_\downarrow$
are suppressed. One may note that when we pass from P to AP alignment, the major
spin up peak has
its width broadened.
This can also be seen by calculating the width of this peak:
$\G_\uparrow=\G_\uparrow^L+\G_\uparrow^R=2\G_0(1-|p|)$ and
$\G_\uparrow=\G_\uparrow^L+\G_\uparrow^R=2\G_0$ for P and AP, respectively. 
In particular, this broadening effect makes the total antiparallel current
slightly higher 
than the parallel current ($I^P<I^{AP}$), and this results into a negative
magnetoresistance, as we
will see next.
 
\begin{figure}
\par
\begin{center}
\epsfig{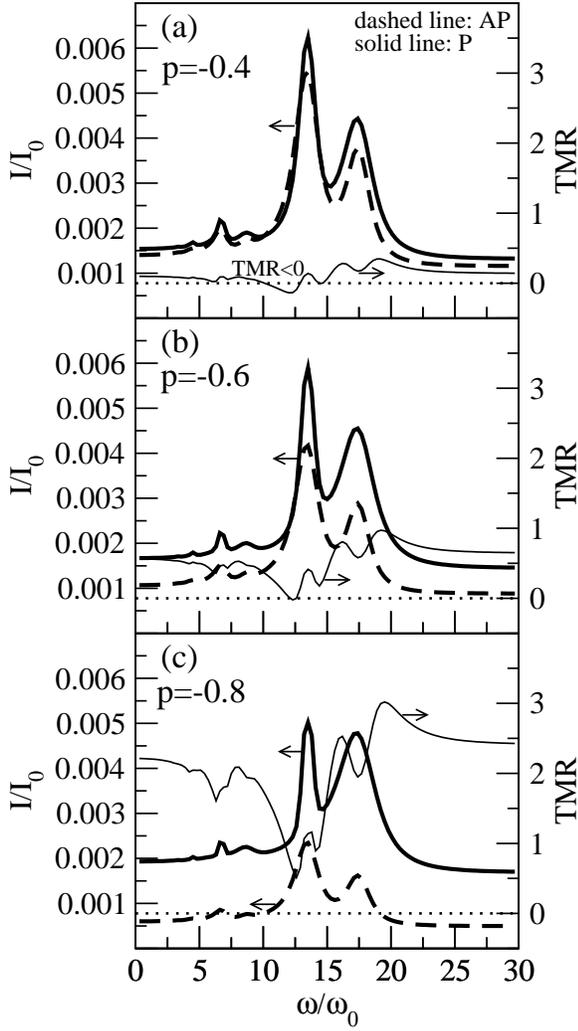}
\end{center}
\caption{Total current ($I_\uparrow+I_\downarrow$) in both P and AP alignment
and TMR as a function
of the
gate frequency. From panels (a) to (c) the left and right leads polarization are
enlarged ($p=-0.4$,
$-0.6$, $-0.8$).
The highest current peak is predominantly due to spin up transport while the
second highest peak
is more spin down like. The TMR tends to be suppressed or amplified around the
spin up or spin down
peaks, respectively. In particular, the TMR attains negative values for $p=-0.4$
in a short gate
frequency range,
due to a photon-assisted inversion of the spin valve effect ($I^P<I^{AP}$).
Units: $I_0=e\G_0/\hbar$
and $\w_0=\G_0/\hbar$.
Parameters: $\e_0=-15\G_0$, $E_Z=4\G_0$, $\D=5\G_0$, $\mu_L=1\G_0$,
$\mu_R=0$.} \label{fig6}
\end{figure}

\begin{figure}[h]
\begin{center}
\epsfig{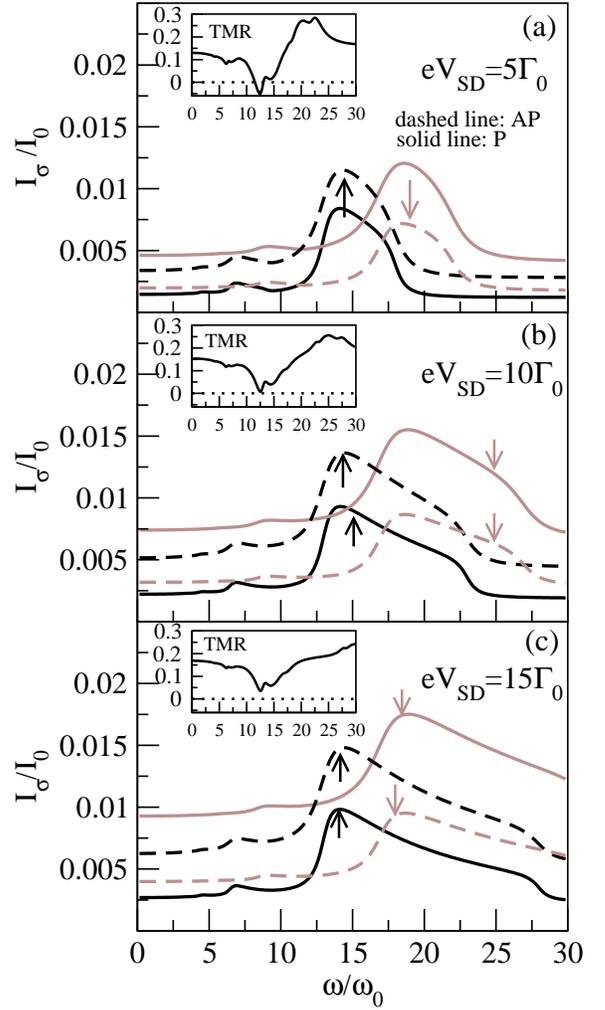}
\end{center}
\caption{(color online) Spin resolved currents against gate frequency for both
parallel and
antiparallel configurations (black lines for spin up and gray lines for spin
down). From panels
(a) to (c) we increase the bias voltage [(a) $eV_{SD}=5\G_0$, (b)
$eV_{SD}=10\G_0$, (c)
$eV_{SD}=15\G_0$], thus
enlarging the
conduction window [dashed lines in Fig. \ref{fig3}]. Fig. \ref{fig7}(a) is
similar to Fig.
\ref{fig5}(a), except by the width of
the peaks that are slightly enlarged by the CW. When $eV_{SD}$ increases even
further
[panels (b)-(c)]
the peaks turn even more
broaden. In the insets we show the TMR for each bias voltage. For
$eV_{SD}=5\G_0$ the
TMR presents a
negative value around $\w \approx 12.5 \w_0$.
For $eV_{SD}=10\G_0$ and $eV_{SD}=15\G_0$ the TMR becomes positive for all
frequencies.
Units:
$I_0=e\G_0/\hbar$ and $\w_0=\G_0/\hbar$. 
Parameters: $\e_0=-15\G_0$, $E_Z=4\G_0$, $\D=5\G_0$, $\mu_L-\mu_R=eV_{SD}$,
$p=-0.4$.}
\label{fig7}
\end{figure}

In Fig. (\ref{fig6}) we show the total current in both P and AP alignment and
the 
tunnel magnetoresistance, defined according to 
\begin{equation}
 TMR=\frac{I^P-I^{AP}}{I^{AP}},
\end{equation}
where $I^{P/AP}=I^{P/AP}_\uparrow+I^{P/AP}_\downarrow$. The highest current peak
is related to the resonance of the channel $\e_\uparrow^{(1)}$ with the 
CW. The second highest peak is due to spin down resonance, $\e_\downarrow^{(1)}
\approx \mu_R$.
The spin valve effect is clearly seen for almost all frequencies, i.e.,
$I^P>I^{AP}$. However, for
frequencies
around 10$\w_0$, we observe $I^P<I^{AP}$ for $p=-0.4$ [Fig. \ref{fig6}(a)].
This photon-assisted opposite spin valve effect is related to the broadening of
the spin up peak
discussed in Fig. (\ref{fig5}),
when the lead polarizations rotate from P to AP alignment. This feature is
reflected in the
TMR, which acquires negative values. Notice that near the position of the spin
up peak 
the TMR is fully suppressed, while near to the peak of the 
spin down current it is enlarged for all values of $p$. This 
indicates that the magnetoresistance is mainly dominated by spin down transport.

Since from the experimental point of view the quantities $V_{SD}$, $\e_d$ and
$E_Z$
are relative {easily} tunable experimental parameters, we
{analyze} how these
quantities
affect the present results, {aiming to} providing a better guide for
future experimental
realizations.
For instance, {using the same set of parameters
adopted previously [$p_L=-0.4$, $p_R=-0.4$(P), $p_R=+0.4$(AP), $\e_0=-15\G_0$,
$\D=5\G_0$,
$E_Z=4\G_0$], in}  Fig. (\ref{fig7}) {we} show how the spin resolved
currents evolve
when the source-drain
voltage increases in both parallel and antiparallel configurations. The bias
voltages considered are (a) $eV_{SD}=5\G_0$, (b) $10\G_0$ and (c) $15\G_0$.
These
values were
also indicated by dashed lines in Fig. (\ref{fig3}). {By} comparing Fig.
\ref{fig7}(a)
to Fig. \ref{fig5}(a) {one can note} that the peaks are broadened as $V_{SD}$
{increases}. This
feature can be understood looking at
the different conduction windows (dashed lines) in Fig. (\ref{fig3}). As
$\mu_L-\mu_R$
becomes larger, the frequency range in which the photon-assisted channels
remains inside
the CW {becomes} wider. In the insets of Fig. (\ref{fig7}) we show the
TMR
against frequency for each $V_{SD}$ considered. {Notice that for
$eV_{SD}=10\Gamma_0$ and
$eV_{SD}=15\Gamma_0$ [panels (b) and (c), respectively] the TMR is positive for
all frequencies.}

\begin{figure}[h]
\centerline{\resizebox{3.5in}{!}{
\includegraphics{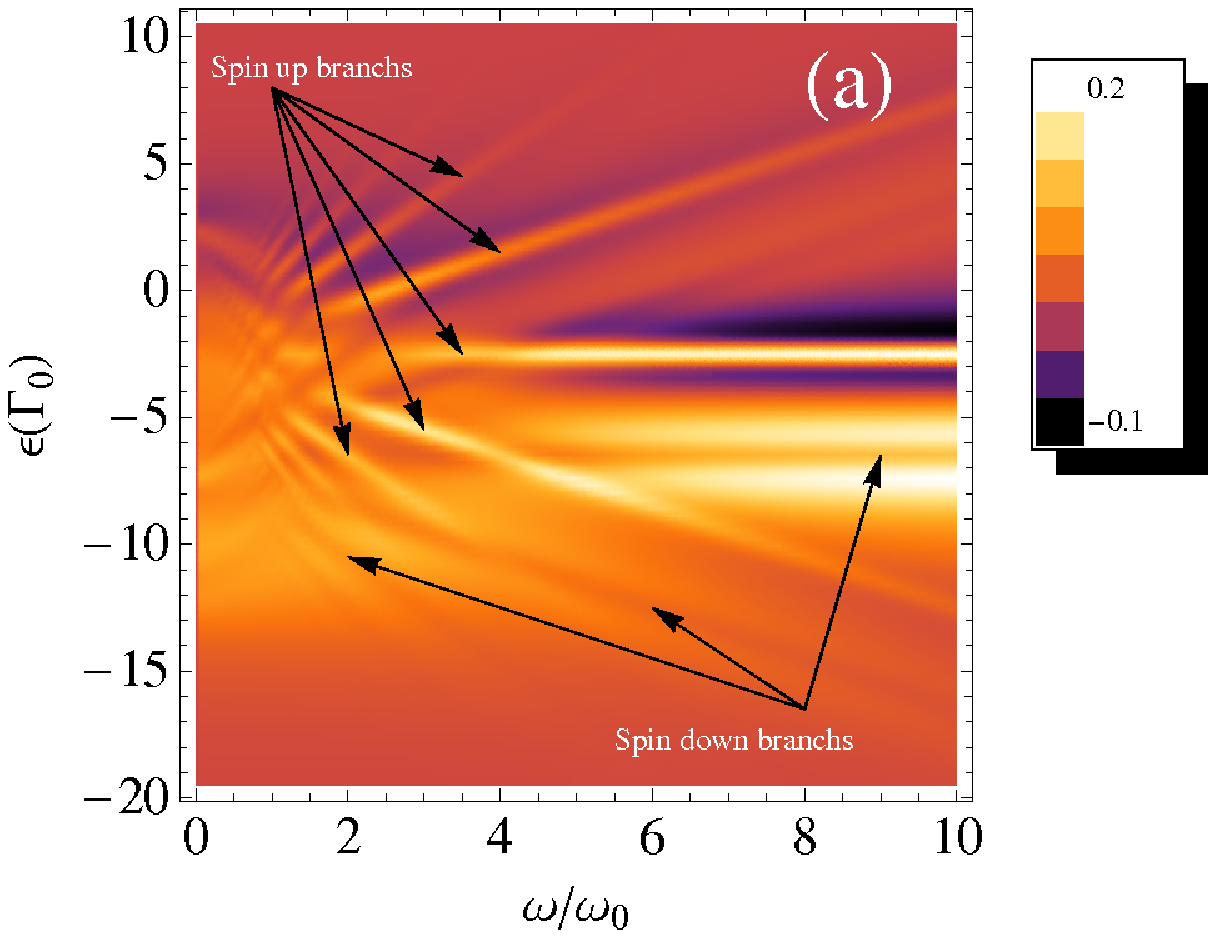}}}
\centerline{\resizebox{3.5in}{!}{
\includegraphics{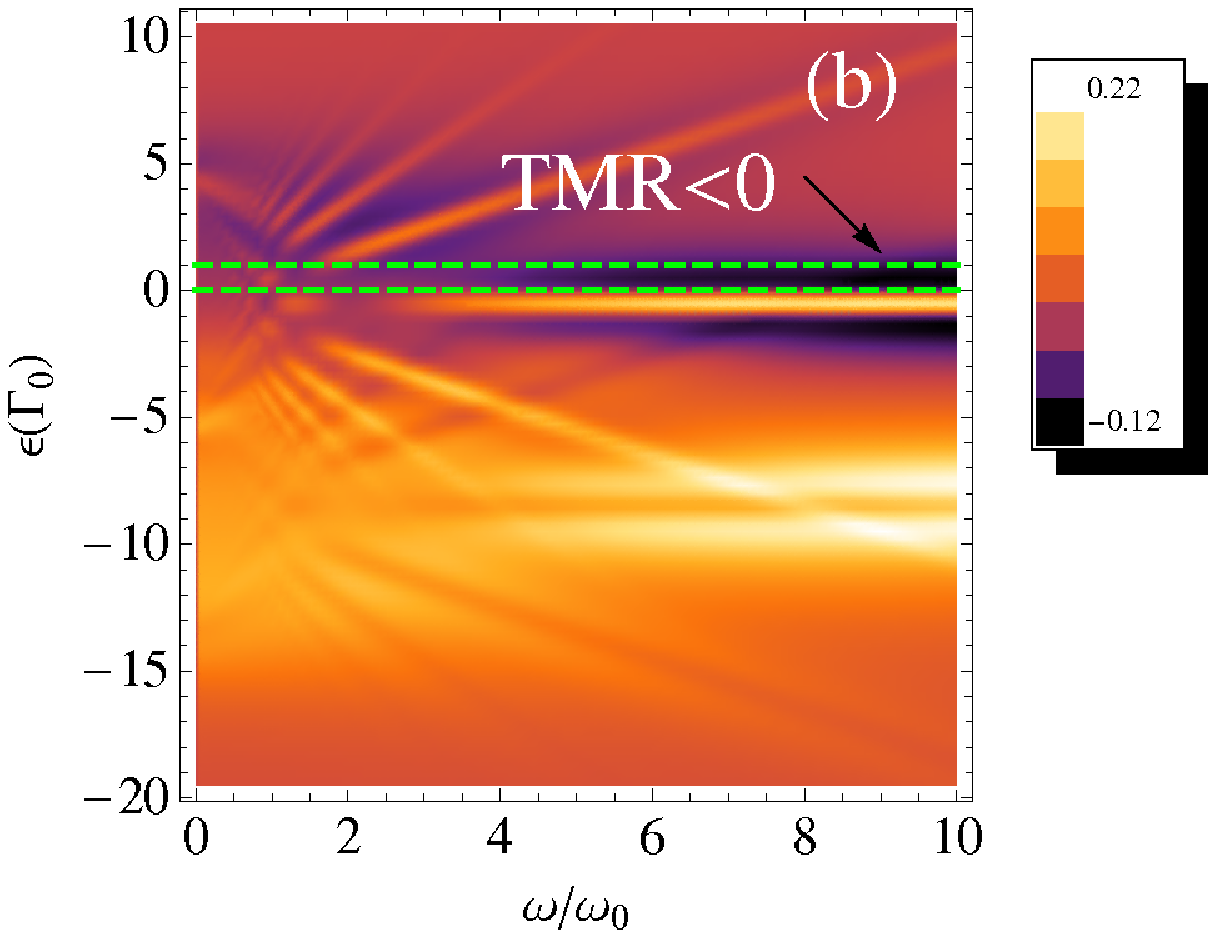}}}
\caption{(color online) Two-dimensional map of the difference $\D T=T^P-T^{AP}$
against
 frequency $\w$ and $\e$ for
two different Zeeman splitting: (a) $E_Z=4 \G_0$ and (b) $E_Z=8 \G_0$. The dark
regions indicate
$T^{AP}>T^P$ which
results in negative TMR. As $E_Z$ increases [from panel (a) to (b)] the dark
ares enlarge. This is
so because
the spin up and down channels become apart from each other, allowing to have
$T^{AP}=T_\uparrow^{AP}+T_\downarrow^{AP} \approx T_\uparrow^{AP} > T^P =
T_\uparrow^{P}+T_\downarrow^{P} \approx T_\uparrow^{P}$
(broadening effect) around the spin up channels. By playing with the set of
parameters one can
place 
the conduction window of the system in the {darker} region of the map,
which results into a
more negative TMR.
Units: $I_0=e\G_0/\hbar$ and $\w_0=\G_0/\hbar$. 
Parameters: $\e_0=-4.5\G_0$, $\D=5\G_0$, $\mu_L=1\G_0$, $\mu_R=0$,
$p=-0.4$.}\label{fig8}
\end{figure}

\begin{figure}
\par
\begin{center}
\epsfig{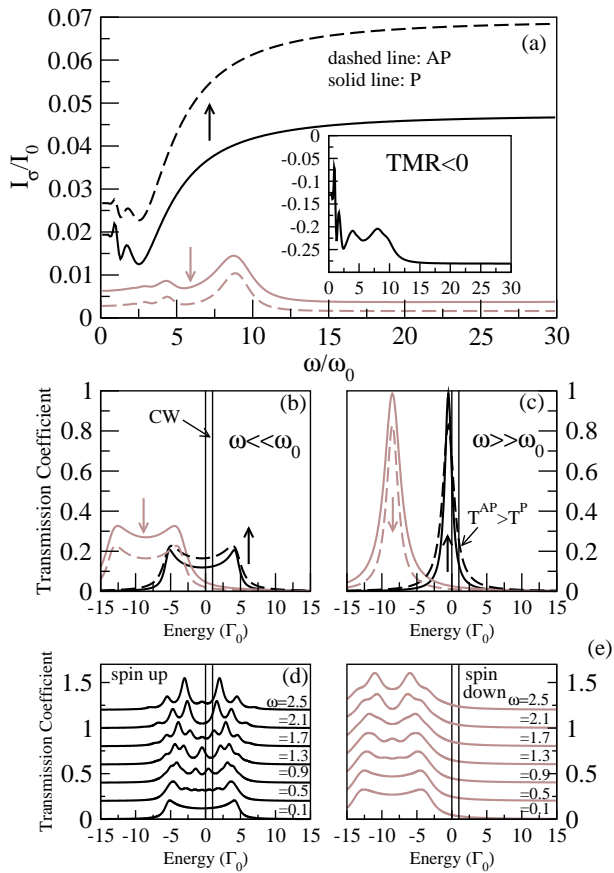}
\end{center}
\caption{(color online) (a) Spin resolved currents in both P and AP alignments
against gate
frequency for the conduction window
drawn in Fig. \ref{fig8}(b). Similar to previous figures, the spin down
component reveals peaks at
some particular frequencies due to the
matches $\e_\s^{(n)} = \mu_R$. In contrast, the spin up current oscillates 
and then it tends to a relatively large saturation value for large $\w$. In the
inset of Fig.
\ref{fig9}(a) we show
the TMR which acquires negative values in accordance to the CW in Fig.
\ref{fig8}(b). Panels (b)-(c)
present the spin resolved transmission coefficient for (b) small and (c) large
$\w$ in both P and AP
configurations.
The thin vertical lines denote the border of the CW. In particular, observe in
panel (c) that
inside the CW we have $T_\uparrow^{AP}>T_\uparrow^{P}$. Since
$T_\downarrow^{P/AP}$ is too small
at the CW, the total transmission coefficient $T^{P/AP}$ is given essentially by
$T_\uparrow^{P/AP}$, so
$T^{AP}=T^{AP}_\uparrow+T^{AP}_\downarrow \approx T^{AP}_\uparrow >
T^{P}=T^{P}_\uparrow+T^{P}_\downarrow \approx T^{P}_\uparrow$ around
the peak. This results in a negative TMR. In panels (d)-(e) we plot (d)
$T_\uparrow^P$ and (e)
$T_\downarrow^P$
for different values of $\w$. The curves were vertically displaced for clarity.
As $\w$
increases additional peaks
emerge mainly in the interval $\e_\s^0 \pm \D = \e_\s^0 \pm 5 \G_0$.
More specifically, in the spin up case the appearance and suppression of these
additional
peaks inside the CW gives rise to the oscillatory patter of $I_\uparrow$ in the
low frequency limit.
Units: $I_0=e\G_0/\hbar$ and $\w_0=\G_0/\hbar$. 
Parameters: $\e_0=-4.5\G_0$, $E_Z=8\G_0$, $\D=5\G_0$, $\mu_L=1\G_0$, $\mu_R=0$,
$p=-0.4$.}
\label{fig9}
\end{figure}

In order to obtain a clear picture on the set of parameters needed to obtain a
negative TMR,
{in Fig.~(\ref{fig8}) we plot in a color map of the total transmission
coefficient
difference, $\D T=T^P-T^{AP}$, against gate frequency (horizontal axis)
and energy $\e$ (vertical axis)}, for two different Zeeman splitting
{energies}: (a)
$E_Z=4\G_0$ and (b) $E_Z=8\G_0$. {Notice}
the appearance of negative values (dark regions) around the spin up peaks, when
the up and down
branches are not overlapping. {The overlap can be avoided by controlling
the Zeeman
splitting. Observe that for $E_Z=8 \G_0$ [panel (b)]  the}
spin up and spin down branches are far enough from each other to ensure
relatively large dark areas.
This effect can be understood in terms of the broadening of the photon-assisted
peaks. In the P
configuration, the spin down peaks at the transmission coefficient are
{broader} than the
spin up ones, as can be seen by the rates
$\G_\downarrow=\G_\downarrow^L+\G_\downarrow^R=2\G_0(1+|p|)=2.8\G_0$ 
and $\G_\uparrow=2\G_0(1-|p|)=1.2 \G_0$ ($|p|=0.4$). So when these peaks
coincide at the same energy
($E_Z=0$) the spin down peaks {lie} above the spin up ones, thus making
the total
transmission
coefficient dominated by the spin down component. When the magnetic alignment is
rotated from
P to AP, the {width of the spin up peak increases and spin down
diminishes, becoming both}
$2 \G_0$. This {facilitates} $T_\uparrow^{AP} > T_\uparrow^{P}$ and 
$T_\downarrow^{AP} <
T_\downarrow^{P}$ nearby each transmission peak.
 If the peaks are far enough from each other ($E_Z\neq 0$) it is possible to
{obtain}
$T=T_\uparrow+T_\downarrow \approx T_\uparrow$ 
around the up peaks, leading to $T^{AP} \approx T_\uparrow^{AP}
> T^P \approx T_\uparrow^{P}$
[see Fig.~\ref{fig9}(c) for clarity]. This is the main {condition for} a
negative
TMR.\cite{commentmajorityup}
{ For instance, by tunning the set of parameters such that the CW
lies  on the darker area of the map, the TMR becomes negative. By increasing the
CW, 
we can eventually cover an energy range with more bright than dark areas,
resulting in a
positive TMR.} 

In Fig.~(\ref{fig9}) we show the current {vs.} $\w$ obtained for the
conduction window set as
in Fig.~\ref{fig8}(b)[dashed (green) lines] Since the CW is dominated by a dark
region
($T^P<T^{AP}$) we
expect $TMR<0$. The spin up and spin down currents in Fig.~\ref{fig9}(a) reveal
contrasting
behavior. While the spin down current shows the same behavior already seen in
previous results
[e.g., Fig.~(\ref{fig5})], the spin up current oscillates and then it increases
to a higher
saturation value. This higher value of $I_\uparrow$ in the large frequency limit
can
be understood by looking at the transmission coefficient in panels \ref{fig9}(b)
{and} (c).
Spin up and spin down transmissions coefficients are drawn in both P and AP
configurations. The
vertical thin solid lines give the border of the CW. Comparing the amplitude of
$T_\uparrow$ inside
the CW for both $\w \ll \w_0$ and $\w \gg \w_0$, we observe that $T_\uparrow$
becomes
{amplified for larger frequencies, which results into higher} spin up
current
in this limit. {Conversely}, the spin down transmission coefficient is
slightly higher in the
CW for $\w \ll \w_0$, {resulting in a spin down} current a bit higher in
the low frequency
limit, compared to its value in the high frequency {regime}.

To better understand the oscillatory structure found in the spin up current, we
plot in
Fig.~\ref{fig9}(d) the spin up parallel transmission coefficient $T_\uparrow^P$
vs. energy for
{various} $\w$. As $\w$ increases the transmission coefficient 
develops a variety of peaks mainly in the range between $\e_\uparrow^0-5\G_0$
and
$\e_\uparrow^0+5\G_0$ ($\e^0_\uparrow \pm \D$). 
Since this energy interval contains the CW, some of the photon-assisted
peaks that arise
for increasing $\w$
{appear within the CW and eventually lie outside it
for large
enough $\omega$},
followed by a new peak emerging inside it.
This results in an oscillatory pattern of the current. In panel \ref{fig9}(e) we
show 
$T_\downarrow^P$ {as function of energy for the same values of $\omega$
as in panel (d)}.
Since $\e^0_\downarrow+\D<\mu_R$ the photon-assisted peaks can only cross
the CW for particular frequencies, which gives rise to the peaks observed.
Finally,
in the inset of Fig. \ref{fig9}(a) we plot the TMR, which presents very low
negative values ($\sim
-30$\%). 
This negative TMR is a result of the set of parameters chosen from Fig.
\ref{fig8}.


\section{Conclusion}
\label{conclusions}
We have studied spin polarized transport in a quantum dot attached to
ferromagnetic leads in the
presence
of an oscillating gate voltage $V_g(t)$. A static source-drain bias voltage is
also applied in order
to generate current. 
The oscillating $V_g(t)$ gives rise to photon-assisted transport channels that
allow
electrons to flow through the system. 
Due to a Zeeman splitting of the dot level, the photon-assisted
{contributions to the transport are
distinct for spins up and down, providing an interesting way to obtain current
polarization that can be controlled by gate frequency.}
As the leads polarization is enlarged, with a majority down population in both
leads (P alignment),
the spin down photon-assisted current peak is enhanced, while the spin up peak
is suppressed. {Moreover}, when the relative polarization alignment of
the leads is switched from P to AP,
the width of the main spin up peak of the current is broadened. This additional
broadening effect results in an opposite
spin valve behavior ($I^P<I^{AP}$) for gate frequencies around the spin up
resonance. As a result,
a photon-assisted negative tunnel magnetoresistance is found.

\section*{Acknowledgments}

The authors acknowledge A. P. Jauho and J. M. Villas-B\^oas for valuable
comments and suggestions. This work was supported by the Brazilian agencies
CNPq, CAPES and FAPEMIG.


\begin{thebibliography}{99}

\bibitem{nsw93} N. S. Wingreen, A. P. Jauho, and Y. Meir, Phys.
Rev. B \textbf{48}, 8487 (1993).

\bibitem{ljg90} L. J. Geerligs, V. F. Anderegg, P. A. M. Holweg, J. E. Mooij,
H. Pothier, D. Esteve, C. Urbina, M. H. Devoret, Phys. Rev. Lett. \textbf{64},
2691 (1990).
\bibitem{lpk91} L. P. Kouwenhoven, A. T. Johnson, N. C. van der Vaart, C. J. P.
M. Harmans, C. T. Foxon, 
Phys. Rev. Lett. \textbf{67}, 1626 (1991).

\bibitem{cb94} C. Bruder and H. Schoeller, Phys. Rev. Lett. \textbf{72}, 1076
(1994).
\bibitem{lpk94_1} L. P. Kouwenhoven, S. Jauhar, K. McCormick, D. Dixon, P. L.
McEuen,
Yu. V. Nazarov, N. C. van der Vaart, and C. T. Foxon, Phys. Rev. B \textbf{50},
2019(R) (1994).
\bibitem{lpk94_2} L. P. Kouwenhoven, S. Jauhar, J. Orenstein, P. L. McEuen,
Y. Nagamune, J. Motohisa, and H. Sakaki, Phys. Rev. Lett. \textbf{73}, 3443
(1994).
\bibitem{bjk95} B. J. Keay, S. J. Allen, Jr., J. Gal\'an, J. P. Kaminski, K. L.
Campman, A. C. Gossard, U. Bhattacharya, and M. J. W. Rodwell,
Phys. Rev. Lett. \textbf{75}, 4098 (1995).
\bibitem{gp04} For a review see G. Platero and R. Aguado, Phys. Rep.
\textbf{395}, 1 (2004).  

\bibitem{cas96} C. A. Stafford and N. S. Wingreen, Phys.
Rev. Lett. \textbf{76}, 1916 (1996).

\bibitem{afa09} A. F. Amin, G. Q. Li, A. H. Phillips, and U. Kleinekath\"ofer,
Eur. Phys. J. B \textbf{68}, 103 (2009).

\bibitem{fc09} F. Cavaliere, M. Governale, and J. K\"onig, Phys. Rev. Lett.
\textbf{103}, 136801 (2009).

\bibitem{rz09} R. Zhu and H. Chen, Appl. Phys. Lett. \textbf{95}, 122111 (2009).
\bibitem{ep09} E. Prada, P. San-Jose, and H. Schomerus, Phys. Rev. B
\textbf{80}, 245414 (2009).
\bibitem{gmmw10} G. M. M. Wakker and M. Blaauboer, Phys. Rev. B \textbf{82},
205432 (2010).
\bibitem{map11} M. Alos-Palop and M. Blaauboer, arXiv: 1102.0926 (2011).

\bibitem{js10} J. Splettstoesser, M. Governale, J. K\"onig, and M. B\"uttiker,
Phys. Rev. B \textbf{81}, 165318 (2010).

\bibitem{sk10} S. Kurth, G. Stefanucci, E. Khosravi, C. Verdozzi, and E. K. U.
Gross, Phys. Rev. Lett. \textbf{104}, 236801 (2010).

\bibitem{spintronics} G. A. Prinz, Science \textbf{282}, 1660
(1998); S. A. Wolf, D. D. Awschalom, R. A. Buhrman, J. M.
Daughton, S. von Moln{\'a}r, M. L. Roukes, A. Y. Chtchelkanova,
and D. M. Treger, \emph{ibid}. \textbf{294}, 1488 (2001);
Semiconductor Spintronics and Quantum Computation, edited by
D. Awschalom, D. Loss, and N. Samarth (Springer, Berlin 2002); D.
D. Awschalom and M. E. Flatt{\'e}, Nat. Phys. \textbf{3}, 153 (2007). 

\bibitem{ec05} E. Cota, R. Aguado, and G. Platero, Phys. Rev. Lett. \textbf{94},
107202 (2005).
\bibitem{rs06} R. S\'anchez, E. Cota, R. Aguado, and G. Platero, Phys. Rev. B
\textbf{74}, 035326 (2006); 
R. S\'anchez, E. Cota, R. Aguado, and G. Platero, Physica E \textbf{34}, 405
(2006).

\bibitem{erm02} E. R. Mucciolo, C. Chamon, and C. M. Marcus, Phys. Rev. Lett.
\textbf{89}, 146802 (2002).
\bibitem{ta03} T. Aono, Phys. Rev. B \textbf{67} 155303 (2003).
\bibitem{skw03} S. K. Watson, R. M. Potok, C. M. Marcus, and V. Umansky, Phys.
Rev. Lett. \textbf{91}, 258301 (2003).

\bibitem{fms07_1} F. M. Souza, Phys. Rev. B \textbf{76}, 205315 (2007).
\bibitem{ep08} E. Perfetto, G. Stefanucci, and M. Cini, Phys. Rev. B
\textbf{78}, 155301 (2008).

\bibitem{pt10} P. Trocha, Phys. Rev. B \textbf{82}, 115320 (2010).

\bibitem{fms07_2} F. M. Souza, S. A. Le{\~a}o, R. M. Gester, and A. P. Jauho,
Phys. Rev. B \textbf{76}, 125318 (2007).
\bibitem{fms09} F. M. Souza and J. A. Gomez, Phys. Status Solid B \textbf{246},
431 (2009).


\bibitem{rh08} R. Hanson and D. D. Awschalom, Nature \textbf{453}, 1043 (2008).

\bibitem{jme04} J. M. Elzerman, R. Hanson, L. H. W. van Beveren, B. Witkamp, L.
M.
K. Vandersypen, and L. P. Kouwenhoven, Nature \textbf{430}, 431
(2004).
\bibitem{rh05} R. Hanson, L. H. W. van Beveren, I. T. Vink, J. M. Elzerman, W.
J. M. Naber, F. H. L. Koppens, L. P. Kouwenhoven, and L. M. K. Vandersypen,
Phys. Rev. Lett. \textbf{94}, 196802 (2005).
\bibitem{acj05} A. C. Johnson, J. R. Petta, J. M. Taylor, A. Yacoby, M. D.
Lukin, C. M. Marcus, M. P. Hanson, and A. C. Gossard,
Nature \textbf{435} 925 (2005).
\bibitem{jrp05} J. R. Petta, A. C. Johnson, J. M. Taylor, E. A. Laird, A.
Yacoby, M. D. Lukin, C. M. Marcus, M. P. Hanson, A. C. Gossard,
Science \textbf{309}, 2180 (2005).
\bibitem{fhlk06} F. H. L. Koppens, C. Buizert, K. J. Tielrooij, I.
T. Vink, K. C. Nowack, T. Meunier, L. P. Kouwenhoven, and L. M. K.
Vandersypen, Nature \textbf{442}, 766 (2006). 
\bibitem{cb10} C. Barthel, J. Medford, C. M. Marcus, M. P. Hanson, and A. C.
Gossard, Phys. Rev. Lett. \textbf{105}, 266808 (2010).

\bibitem{hh96} H. Haug and A. P. Jauho, Quantum Kinetics in
Transport and Optics of Semiconductors, Springer Solid-State
Sciences \textbf{123} (1996).

\bibitem{apj94} In the present calculation we follow the formulation developed
in 
A. P. Jauho, N. S. Wingreen, and Y. Meir, Phys. Rev. B \textbf{50}, 5528 (1994).

\bibitem{wr01} W. Rudzi{\'n}ski and J. Barna{\'s}, Phys. Rev. B
\textbf{64}, 85318 (2001).

\bibitem{fms04} F. M. Souza, J. C. Egues, and A. P. Jauho, Braz. J. Phys.
\textbf{34}, 565 (2004).

\bibitem{commentnoncollinear} {In the present work we consider
only the P and AP alignments.
We expect that for more general non-collinear magnetizations the present
currents will vary
monotonically from their parallel ($\phi=0$) to their antiparallel ($\phi=\pi$)
values. In the
adiabatic-pumping regime it was found (see, for instance, Ref.
[\onlinecite{js08}]) a monotonic suppression of the pumped charge as the 
angle between the magnetizations of the left and right leads {is
continuously varied} from
$\phi=0$ to $\phi=\pi$.}

\bibitem{js08} J. Splettstoesser, M. Governale, and J. K\"onig, Phys. Rev. B
\textbf{77} 195320 (2008).

\bibitem{commentreviewplatero} An alternative but equivalent version of this
equation was
derived in Ref. [\onlinecite{gp04}], where the distinction between this result
and the
Tien-Gordon formula was discussed.

\bibitem{dgg98_1}D. G.-Gordon, H. Shtrikman, D. Mahalu, D.
A.-Magder, U. Meirav, M. A. Kastner, Nature \textbf{391}, 156 (1998).
\bibitem{dgg98_2} D. Goldhaber-Gordon, J. G{\"o}res, M. A. Kastner, H.
Shtrikman, D. Mahalu, and U. Meirav,
Phys. Rev. Lett. \textbf{81}, 5225 (1998).
\bibitem{fs99} F. Simmel, R. H. Blick, J. P. Kotthaus, W. Wegscheider, and M.
Bichler, Phys. Rev. Lett.
\textbf{83}, 804 (1999). 

\bibitem{rh03} R. Hanson, B. Witkamp, L. M. K. Vandersypen, L. H. Willems van
Beveren, J. M. Elzerman, and L. P. Kouwenhoven,
Phys. Rev. Lett. \textbf{91}, 196802 (2003).

\bibitem{thzelectronics} Significant improvements on THz electronics have been
recently achieved. See for instance,
T. W. Crowe, W. L. Bishop, D. W. Porterfield, J. L. Hesler, R. M. Weikle, IEEE
J. Solid-State Circuits, \textbf{40} 2104 (2005).

%
\bibitem{sami} S.~Amasha (private communication).
\bibitem{km07} K. MacLean, S. Amasha, I. P. Radu, D. M. Zumb\"uhl, M. A.
Kastner, M. P. Hanson, and A. C. Gossard, Phys. Rev. Lett. \textbf{98}, 036802
(2007). 
%
\bibitem{he10} H. Eisele, S. P. Khanna, and E. H. Linfield, Appl. Phys. Lett.
\textbf{96} 072101 (2010).

\bibitem{commentotherwork} Two split peaks in a pumping current, one peak for
each spin component,
were also found by Cota \emph{et al.} in Ref. [\onlinecite{ec05}] for a double
dot system in the presence of magnetic field.

\bibitem{commentmajorityup} {In the case of majority up
population in both leads in the P alignment, 
we have the opposite behavior seen in Fig. (\ref{fig8}), i.e., the maxima become
minima and vice-versa. So the
dark areas in the plot ($T^P<T^{AP}$) arise around the spin down branches,
instead of the spin up ones.}
\end{thebibliography}
\end{document}